\documentclass[twocolumn,aps,floats,floatfix,prl,nofootinbib,superscriptaddress,tightenlines]{revtex4}

\usepackage{graphicx}
\usepackage{bm}
\usepackage{epsfig}
\usepackage{pstricks}
\usepackage{amsmath}
\usepackage{hyperref}
\usepackage{color}

\newcommand{\beq}{\begin{equation}}
\newcommand{\eeq}{\end{equation}}
\newcommand{\bea}{\begin{eqnarray}}
\newcommand{\eea}{\end{eqnarray}}
\newcommand{\OMIT}[1]{}
\newcommand{\bF}{{\bf F}}
\newcommand{\bx}{{\bf x}}
\newcommand{\bv}{{\bf v}}

\begin{document}

\title{Finite Size Corrections to the Radiation Reaction Force in Classical Electrodynamics}

\def\addCMU{Department of Physics, Carnegie Mellon University, Pittsburgh PA  15213, USA}
\def\addMaryland{Maryland Center for Fundamental Physics, Department of Physics,
University of Maryland, College Park, MD 20742 USA}
\def\addModeling{Center for Scientific Computation and Mathematical
Modeling, University of Maryland, College Park, MD 20742 USA}
\def\addPitt{Department of Physics and Astronomy, University of
  Pittsburgh, PA 15260, USA} 

\author{Chad R. Galley}
\affiliation{\addMaryland}\affiliation{\addModeling}
\author{Adam K. Leibovich}
\affiliation{\addPitt} 
\author{Ira Z. Rothstein}
\affiliation{\addCMU}

\begin{abstract}

We introduce an effective field theory approach that describes the motion of  finite size
objects under the influence of electromagnetic fields. We prove that leading order effects due to the finite radius $R$ of a spherically symmetric charge is order $R^2$ rather than order $R$ in any physical model, as widely claimed in the literature. This scaling arises as a consequence of Poincar\'e and gauge symmetries, which can be shown to exclude linear corrections. 
We use the formalism to calculate the leading order finite size  correction to
the Abraham-Lorentz-Dirac force.

\end{abstract}

\maketitle 

Radiation reaction in classical electrodynamics has a long and interesting history.  It was first solved for a non-relativistic point particle  by Abraham and Lorentz.  The relativistically 
covariant result was derived  some years later by Dirac\footnote{We use naturalized Heaviside-Lorentz units with $c=1$.} and is given by the ALD equation
\bea
m\ddot x^\mu &=& F_{\rm ext}^\mu  +  F_{\rm self}^\mu    \nonumber\\
&=& F_{\rm ext}^\mu  + \frac23 \frac{e^2}{4\pi} \left(\dddot x^\mu - \dddot x^\nu \dot x_\nu \dot x^\mu\right) 
\label{aldeq_1}
\eea
where $F_{\rm ext}^\mu$ is the external force accelerating the charge.

Corrections to the point particle limit have been investigated in the literature in the past.
However, as we will show in this letter, these corrections have yet to be treated
properly.
Worldline repararmeterization invariance (RPI) restricts the finite size corrections to have
the form
\beq
F_{\rm self}^\mu = \sum_{n=0}^\infty I_n \omega_n^{\mu\nu} \left(\frac{\partial}{\partial\tau}\right)^n \ddot x_\nu
\eeq
where the $I_n$ are coefficients that depend on the charge distribution of the object. The tensors $\omega_n^{\mu\nu}$ ensure orthogonality of the equation with $\dot x_\mu$ and will in general depend on proper time derivatives of $x^\mu$ leading to nonlinear powers of the acceleration and its derivatives.  For a given charge distribution, the above equation can be expanded in a power series in $R$, the size of the object,
\beq\label{selfFexp}
F_{\rm self}^\mu = F_0^\mu + RF_1^\mu  + R^2 F_2^\mu + \cdots + R^n F_n^\mu + \cdots .
\eeq

A calculation of the non-relativistic self force, neglecting the nonlinear powers of the acceleration and its derivatives, can be found in the classic text by Jackson \cite{jackson} where the force for an arbitrary spherical charge distribution is given by
\beq
{\bF}_{\rm self} = -\frac23 \sum_{n=0}^\infty \frac{(-1)^n}{n!} I_n\left(\frac{\partial}{\partial t}\right)^n\ddot{\bx} ,
\eeq
with
\beq
I_n = \int d^3x'\int d^3 x \, \rho(\bx') |\bx - \bx'|^{n-1} \rho(\bx).
\eeq
For a charged shell, the sum can be done explicitly giving
\bea\label{wrong}
\bF_{\rm self} &=& \frac23 \frac{e^2}{2R^2}\bv(t-2R)\qquad[\bv(0) = 0]\nonumber\\
&=& \frac23 e^2 \left(- \frac1R \ddot{\bx} +  \dddot{\bx} -\frac23 R \bx^{(4)}+ \cdots \right)  .
\eea
The first term is the usual infinite self-energy of the point particle which renormalizes the mass, the second term is the ALD radiation reaction, and the third is the first finite size correction proportional to the size $R$ of the object.
This result was covariantized in Ref.~\cite{Caldirola, Yaghjian, Rohrlich2} 
for relativistic motions, which exhibits a similar expansion in powers of $R$.
In this letter, we utilize the worldline formalism developed in Ref.~\cite{Goldberger:2004jt} to show that  symmetries prohibit linear corrections in Eq.~(\ref{wrong}) and its generalization in Ref.~\cite{Caldirola, Yaghjian, Rohrlich2}, or more generally in Eq.~(\ref{selfFexp}), for a spherically symmetric charge distribution.

The method used in this paper to calculate the radiation reaction is based on an effective field theory (EFT) approach  developed in Ref.~\cite{Goldberger:2004jt}.  The worldline action for a point particle
\beq\label{ppaction}
S^{(0)} = -m \int d\tau + e \int d\tau\, v^\mu A_\mu
\eeq
 is systematically augmented by interactions with increasing powers of the radius.  All terms consistent with the symmetries (i.e., Poincar\'e, gauge and RPI) are included.  The additional terms,  in the augmented point-particle action, have coefficients that encapsulate the structure of the charged object, including finite size effects, susceptibility, etc. Given some model for the charged object these coefficients can be fixed by a matching calculation.

To calculate the equations of motion we apply the Euler-Lagrange equations
to the highly non-local effective action obtained by integrating out the electromagnetic field,
\beq
\exp\left\{i {\cal S}^{\rm eff}[J]\right\} = \int {\cal D}A_\mu \exp\left\{i S_{\rm EM} + i S_{pp}\right\}
\eeq
where $S_{\rm EM}$ is the electromagnetic action and $S_{pp}$ is the point particle action.  
In practice, we utilize a diagrammatic approach as in quantum field theory.  Classical physics results from calculating the ``tree level" Feynman diagrams.  
The ``usual" Feynman rules for scattering processes (found in quantum field theory textbooks) must be slightly modified to incorporate the retarded boundary conditions on the field at the level of the action.  For the calculations presented below, the modifications amount to using the retarded propagator for the photon line, which can be justified using the ``in-in" formalism \cite{Galley:2009px}.  
We begin by demonstrating how to obtain the ALD equation from the point particle action.

\begin{figure}[t]
\includegraphics[width=2in]{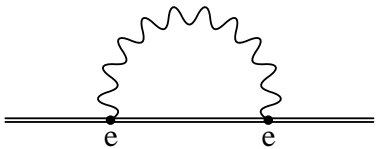}
\caption{The Feynman diagram giving the effective action, which when varied leads to the ALD radiation reaction equation, Eq.~(\ref{aldeq_1}). The wiggly line represents a retarded photon and the double line indicates the charge's worldline, which is to be viewed as an external source coupling to the photon. The dot represents the coupling of the photon to the charge.}
\label{aldfig}
\end{figure}

The leading order radiation reaction (i.e., ALD) equation, Eq.~(\ref{aldeq_1}), 
follows by calculating the Feynman diagram shown in
 Fig.~\ref{aldfig}.
In this figure the double line represents the worldline of the particle, the dots are insertions of the $v\cdot A$ operator\footnote{For historical reasons the term ``operator''  is used despite the obvious classical nature of this problem.}, and the wiggly line represents the propagator for the vector field $A^\mu$.  
The result of evaluating Fig.~\ref{aldfig} is
\begin{align}
{\rm Fig. \,} \ref{aldfig}  = i {\cal S}^{\rm eff}_0 = {} &  (i e )^2 \int d\lambda\, d\lambda' \, v^\alpha (\lambda) v^\beta (\lambda')  \nonumber \\
	& \times  \big\langle A_\alpha \big( x(\lambda) \big) A_\beta \big( x(\lambda') \big) \big\rangle_{ret}\,.
\label{ald1}
\end{align}
The two point function above is replaced by the retarded propagator in the Lorentz gauge
\bea
i \big\langle A_\alpha (x) A_\beta (y) \big\rangle_{ret} &=& D^{ret} _{\alpha \beta} (x-y) \\
 &=& \frac{g_{\alpha\beta}}{2\pi}\theta(x^0-y^0)\delta[(x-y)^2] \nonumber
\eea
and the effective action ${\cal S}^{\rm eff}_0$ is varied with respect to the {\it most recent} time parameter\footnote{This restriction follows from simple physical arguments and can be proven using the in-in formalism.} giving
\bea
F_0^\mu (\tau)  &=& - e^2 ( g^{\mu\alpha} v^\beta - g^{\mu \beta} v^\alpha ) \label{ald2} \\
&&\times \int d\lambda \,  v_\alpha (\lambda) \partial_\beta D_{ret} \big( x(\tau) - x(\lambda) \big)  \nonumber
\eea
where $\tau$ is the proper time of the charge.
This equation diverges when $\lambda\to\tau$.  To proceed further we follow a trick introduced by Coleman \cite{Coleman:1969xz}.  Define $z^2 = [x(\tau)-x(\lambda)]^2$
and use $z$ as the new affine parameter for the integral in Eq.~(\ref{ald2}).  For simplicity, we set $\tau =0$.  The derivative acting on the retarded propagator can be written as
\beq
\partial_\beta = \frac{x_\beta(z)-x_\beta(0)}{z}\frac{d}{dz}
\eeq
so that, after integrating by parts, Eq.~(\ref{ald2}) becomes
\bea
F_0^\mu &=& \frac{e^2}{2\pi} ( g^{\mu\alpha} v^\beta - g^{\mu \beta} v^\alpha )\\
&&\times \int_0^\infty dz\, \delta(z^2)\frac{d}{dz}\left[v_\alpha (z)\frac{x_\beta(z)-x_\beta(0)}z\right] .\nonumber
\label{ald3}
\eea
Expanding the expression inside the square brackets about $z=0$ gives 
\bea\label{zexp}
&& {\hskip-0.1in} \frac{d}{dz} \left[v_\alpha (z)\frac{x_\beta(z)-x_\beta(0)}z\right]  \nonumber\\
&& =  \left(x''_\alpha x'_\beta + \frac12x'_\alpha x''_\beta\right)  + z\left(x_\alpha^{(3)} x'_\beta + x_\alpha''x_\beta'' + \frac13 x_\beta^{(3)}x'_\alpha\right) \nonumber \\
&& ~~~ +O(z^2)
\eea
where the primes denote differentiation with respect to $z$.  The order $z^0$ term above yields a divergence that, after performing the contractions, can be written as
\beq
 \left[\frac{e^2}{4\pi} \int dz\,\delta(z^2) \right] a^\mu.
\eeq
This can be absorbed into a redefinition of the bare mass term in the point particle action and is the usual infinite self energy of a charged point particle.

The order $z$ term in Eq.~(\ref{zexp}) contributes a finite piece, which after the contractions is
\beq
F_0^\mu = \frac23 \frac{e^2}{4\pi} \left(\dddot x^\mu - \dddot x^\nu \dot x_\nu \dot x^\mu\right),
\eeq
the ALD self-force.  Higher orders in $z$ give no contribution.   We are thus able to exactly reproduce the ALD equation using the point particle worldline action.

At this point, we can go beyond the point particle description by including higher order operators in the worldline action.  For a spherical charge distribution, meaning that there are no inherent parameters describing the shape other than the radius in the instantaneous rest frame, 
the first possible operator consistent with Poincar\'e invariance, RPI and gauge symmetry that can be added to the action enters at order $R^2$,
\beq\label{S2}
S^{(2)} = -m \int d\tau + e \int d\tau\, v^\mu A_\mu + C_2 R^2\int d\tau \, O_2,
\eeq
where 
$$O_2 =  \frac{ v_\mu a_\nu F^{\mu\nu} }{ v^2} $$ 
describes an acceleration-induced dipole moment on the charge and $C_2$ is an unknown coefficient at this point.  The factor of $v^2$ is inserted to ensure RPI for a generic affine worldline parameterization.  The numerical value of $C_2$ depends on the shape of the charge distribution and can only be determined from a matching calculation.    

Note that there are no operators consistent with the symmetries at order $R$ for a spherically symmetric charge distribution. Operators that enter at higher order in $R$ come with either more derivatives or more fields.  The operators that are Lorentz invariant that could enter at order $R$ would be
\beq
\frac{v_\mu \dot A^\mu}{\sqrt{v^2}},\quad\frac{a_\mu A^\mu}{\sqrt{v^2}},\quad\sqrt{v^2}\partial_\mu A^\mu,\text{\ \ or\ \ } \sqrt{v^2}A_\mu A^\mu.
\eeq
While these operators are Lorentz invariant, they are not gauge invariant, so they must be excluded.  Thus the first correction enters at order $R^2$.
In addition, if the object is not polarizable then gauge and time-reversal invariance ensures that all higher order corrections enter with even powers of $R$.  If, in an undisturbed state, the charge were not spherically symmetric, then we would need to introduce some auxiliary geometric objects to describe its orientation, which, in turn, would
then allow for new sets of operators.

As an example of a matching calculation, we will determine $C_2$ for the case of a spherical shell of charge.  To accomplish this we  match the result of calculating any observable, for which the operator $O_2$ contributes, using both the worldline action and ``full" electrodynamics.  In this case, it is simplest to calculate the power radiated by a charged spherical shell undergoing harmonic oscillation.  The calculation using electrodynamics was previously done in Ref.~\cite{marengo}.  The power radiated can be obtained from Eq.~(39) of that paper and gives, after expanding for small $R$,
\beq
P_{\rm EM} = \frac{2}{3}\frac{e^2}{4\pi} \left[ \ddot {\bx}^2 -\frac{R^2}3 \dddot{\bx}^2  +\cdots \right].
\eeq
Doing the same calculation using the worldline action is straightforward using standard field theory techniques. The power is given by
\beq
P_{\rm WL} = \frac1{2T}\int \frac{d^3k}{(2\pi)^3} |\tilde J(k)|^2
\eeq
where $\tilde J(k)$ is the Fourier transform of the current.  In our case, 
\begin{widetext}
\beq
 |\tilde J(k)|^2 = \int d\tau\, d\tau' e^{i k\cdot[x(\tau)-x(\tau')]} \left\{e^2 v_\mu(\tau) \varepsilon^\mu  
 v_\nu(\tau') \varepsilon^\nu +e C_2 R^2  \left[v_\mu(\tau) a_\nu(\tau) i k^{[\mu}\varepsilon^{\nu]} 
 v_\alpha(\tau') \varepsilon^\alpha + {\rm h.c.}\right] +  \cdots \right\}  .
\eeq
\end{widetext}
After summing over polarizations and taking the nonrelativistic limit we obtain
\beq
P_{\rm WL} = \frac{2}{3}\frac{e}{4\pi} \left[ e \ddot {\bx}^2 - 2 C_2 R^2 \dddot{\bx}^2  +\cdots \right].
\eeq
By equating the expression for $P_{\rm WL}$ with $P_{\rm EM}$, we find $C_2$ for a charged, spherical shell to be
\beq\label{cshell} 
C_2^{\rm shell} = \frac{e}{6}.
\eeq

\begin{figure}[t]
\includegraphics[width=1.5in]{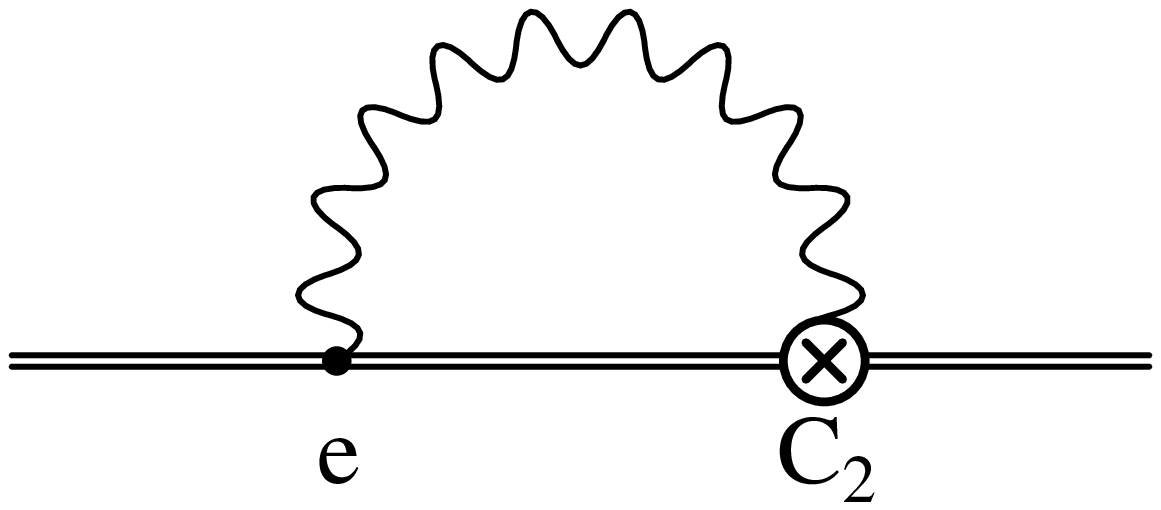}
\includegraphics[width=1.5in]{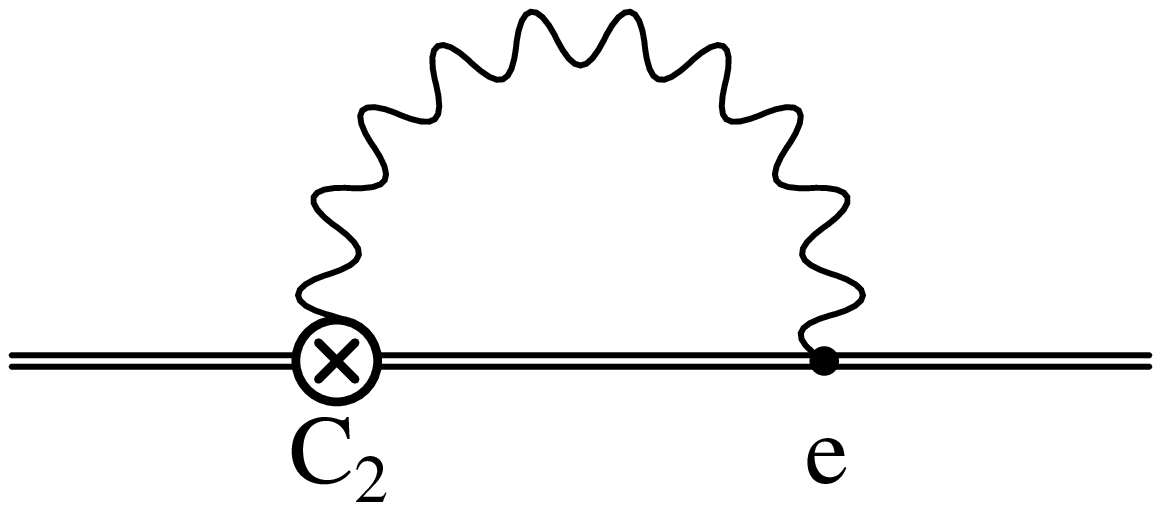}
\caption{Feynman diagrams needed for the first finite size corrections to the ALD equation.  The $\otimes$ represents the coupling of the photon to the operator $O_2$.}
\label{aldplus}
\end{figure}

To calculate the first finite size correction to the ALD equation, Eq.~(\ref{aldeq_1}), we follow the same method used above but instead with the action in Eq.~(\ref{S2}). To include the finite size effects we calculate the two diagrams shown in Fig.~\ref{aldplus}.
Again, the ``in-in" formalism dictates the use of the retarded propagator for the photon line and the variation will be taken only at the most recent time.  After a lengthy but straight-forward calculation, the finite part of the order $R^2$ correction to the ALD equation is
\bea\label{f2}
F_2^\mu &=& \frac43e C_2R^2 \left[\left(x^{(5)\mu} - x^{(5)\nu} \dot x_\nu \dot x^\mu\right) \right.  \label{dipforce} \\
&&\left. + 2 
\dddot x^\rho \dot x_\rho\left( \dddot x^\mu - \dddot x^\nu \dot x_\nu \dot x^\mu\right) - 2  \ddot x^\mu x^{(4)\nu}\dot x_\nu
\right]. \nonumber
\eea
There is also a divergent term that, if regulated by  a high frequency cut-off $\Lambda$, grows linearly with $\Lambda$ 
 and is proportional to 
\beq
e C_2 R^2  \Lambda \left(x^{(4)\mu} - x^{(4)\nu} \dot x_\nu \dot x^\mu\right).
	\label{dipdiv1}
\eeq
This term cannot be absorbed into the particle's mass and charge. On dimensional grounds, the divergence can only be absorbed by including $O(R)$ terms into Eq.~(\ref{S2}), such as $C \int d\tau \, A^2 (\tau)$, which are all gauge-violating. 
However, it is well-known that a cut-off regularization scheme can break gauge-invariance by inducing gauge-violating counter terms into the theory. This happens, for example, with the photon self-energy in quantum electrodynamics (see e.g., \cite{Peskin:1995ev}). Of course, one should regularize the divergences in a manner that preserves the gauge symmetry by using, for example, dimensional regularization which automatically sets all power-divergent integals, such as  Eq.~(\ref{dipdiv1}), to zero. Only logarithmic divergences are physically relevant and, if present,  lead to  a classical renormalization group flow for the coupling constants of the effective theory \cite{Goldberger:2004jt}. 

If we use the matching coefficient in Eq.~(\ref{cshell}) for a spherical shell of charge, then the order $R^2$ correction to the ALD equation in Eq.~(\ref{dipforce}) becomes
\bea\label{f2shell}
F_{2,{\rm shell}}^\mu &=& \frac29e^2 R^2 \left[\left(x^{(5)\mu} - x^{(5)\nu} \dot x_\nu \dot x^\mu\right) \right.\\
&&\left. + 2 
\dddot x^\rho \dot x_\rho\left( \dddot x^\mu - \dddot x^\nu \dot x_\nu \dot x^\mu\right) - 2  \ddot x^\mu x^{(4)\nu}\dot x_\nu
\right]. \nonumber
\eea
To arrive at this result using canonical methods, the calculation of the radiation reaction must be done starting from a completely relativistic theory for the motion of a charged sphere. Such a calculation was explored in the contribution  by Pearle in  Ref.~\cite{teplitz} by calculating in the instantaneous rest frame as proposed by Fermi \cite{fermi}, Wilson \cite{wilson}, Kwal \cite{kwal}, and Rohrlich \cite{rohrlich}.  Pearle did not derive the result in Eq.~(\ref{f2shell}).  However, by applying Eq.~(9.26) from that chapter to a spherical shell of charge one can  obtain Eq.~(\ref{f2shell}).  Furthermore, it is easy to see that the corrections which arise in this methodology  arise with only even powers of $R$ for any spherical charge density, just as the effective theory predicts on symmetry grounds.  

Within the EFT approach this relativistic charged spherical shell model would have vanishing coefficients 
for the operators 
responsible for generating  multipole susceptabilities.
This is consistent with the fact  that the  formalism used in Ref.~\cite{teplitz} does not include the effects of charge mobility.
In the EFT approach the susceptabilities are generated by operators quadratic in the
field strength.  The dipole susceptabilities are accounted for by including two new operators
in the action
\beq
S_{\rm susc}=C_{E} R^3\int d\tau\, F_{\mu\nu}F^{\mu\nu} + C_B R^3 \int d\tau\, v^\mu v^\nu F_{\mu\rho}F^\rho_{\ \nu}.
\label{Lsus}
\eeq
The resulting corrections to the ALD equation arising from the inclusion of these operators will be presented in a forthcoming paper \cite{future}.

To understand the difference between our result in Eq.~(\ref{f2shell}), which includes terms nonlinear in the acceleration and its derivatives, and Eq.~(\ref{wrong}) we can also use the result of Pearle \cite{teplitz}.  If we start from Eq.~(9.26) of that chapter and first take the nonrelativistic limit {\it before} expanding in the size of the object then we obtain Eq.~(\ref{wrong}).  However, the order of limits does not commute and the correct
result will only arise when one takes the small radius limit first.

In summary, we have utilized the worldline effective theory formalism to describe the motion of an extended charge distribution interacting with  an electromagnetic field. We have shown that {\it any} accelerated spherically symmetric charge distribution experiences self-force corrections due to its finite size beginning at order $R^2$ instead of order $R$, as claimed in the literature.   Poincar\'e and gauge symmetries prevent any order $R$ terms from  appearing. Our effective theory description is also a model-independent way to describe charge distributions with non-vanishing susceptabilities by including the operators in Eq.~(\ref{Lsus}). It also possible
to extend the formalism to allow for absorptive effects \cite{GnR2} and
spin \cite{Porto}.

We we would to thank J.~D.~Jackson, P.~M.~Pearle, Ted Jacobson and Walter Goldberger for helpful discussions.  C.R.G.~is supported in part by National Science Foundation grants PHY0801213 and PHY0908457.
A.K.L.~is supported in part by the National Science Foundation under Grant No.~PHY-0546143 and in part  by the Research Corporation.  I.Z.R.~is supported in part by DOE Grants DOE-ER-40682-143 and DEAC02-
6CH03000.


\begin{thebibliography}{99}
\baselineskip 3.0ex 

\bibitem{jackson}
J.~D.~Jackson, Classical Electrodynamics, 3rd ed. (Wiley, New York, 1999).

\bibitem{Caldirola}
	P. Caldirola, Nuovo Cimento {\bf 3}, Suppl.\ 2, 297 (1956).
\bibitem{Yaghjian}
	A.~D.~Yaghjian, Relativistic Dynamics of a Charged Sphere (Springer-Verlag, Berlin, 1992).
\bibitem{Rohrlich2}
	F. Rohrlich, Am.\ J.\ Phys. {\bf 65}, 1051 (1997).

\bibitem{Goldberger:2004jt}
  W.~D.~Goldberger and I.~Z.~Rothstein,
  Phys.\ Rev.\  D {\bf 73}, 104029 (2006)
  [arXiv:hep-th/0409156].

\bibitem{Galley:2009px}
  C.~R.~Galley and M.~Tiglio,
  Phys.\ Rev.\  D {\bf 79}, 124027 (2009)
  [arXiv:0903.1122 [gr-qc]].

\bibitem{Coleman:1969xz}
  S.~Coleman,
{\it  In *Erice 1969, Ettore Majorana School On Subnuclear Phenomena*, New York 1970, 282-327}

\bibitem{marengo}
E.~A.~Marengo and M.~R.~Khodja, Phys.\ Rev.\ E {\bf 74}, 036611 (2006).

\bibitem{Peskin:1995ev}
  M.~E.~Peskin and D.~V.~Schroeder,
{\it  Reading, USA: Addison-Wesley (1995) 842 p}



\bibitem{teplitz}
P.~Pearle in ``Electromagnetism: Paths to Research", Vol.~I, Chapter 7, Ed.~Doris Teplitz, pp.~211-295 (Plenum, New York 1982). 

\bibitem{fermi}
E.~Fermi, Physik Z.\ 23, 340 (1922); Atti Acad.\ Nazl.\ Lincei 31, 184 (1922) ; Atti Acad.\ Nazl.\ Lincei 31, 306 (1922).

\bibitem{wilson}
W.~Wilson Proc.\ R.\ Soc.\ 48, 736 (1936).

\bibitem{kwal}
B.~Kwal, J.\ Phys.\ Radium 10, 103 (1949).

\bibitem{rohrlich}
F.~Rohrlich, Am.\ J.\ Phys.\ 28, 639 (1960).


\bibitem{future}
C.~R.~Galley, A.~K.~Leibovich and I.~Z.~Rothstein, work in progress. 
\bibitem{GnR2}
  W.~D.~Goldberger and I.~Z.~Rothstein,
  Phys.\ Rev.\  D {\bf 73}, 104030 (2006)
  [arXiv:hep-th/0511133].
\bibitem{Porto}
  R.~A.~Porto,
  Phys.\ Rev.\  D {\bf 73}, 104031 (2006)
  [arXiv:gr-qc/0511061], 
  Phys.\ Rev.\  D {\bf 77}, 064026 (2008)
  [arXiv:0710.5150 [hep-th]].
\end{thebibliography}
\end{document}